\documentstyle[12pt,a4]{article}
\def\be{\begin{equation}}
\def\ee{\end{equation}}
\def\barr{\begin{array}}
\def\earr{\end{array}}

\newcommand{\nobody}{\rule{0ex}{1ex}}
\newcommand{\nobodyfrac}{\frac{\nobody}{\nobody}}

\newcommand{\sm}{standard model}
\newcommand{\cm}{center of mass}
\newcommand{\xs}{cross section}
\newcommand{\lc}{linear collider}

\begin{document}
\thispagestyle{empty}
\setcounter{page}{0}

\title{
\vspace{-2cm}
\begin{flushright}
{\large MPI-Ph/94-23\\
LMU-04/94\\
hep-ph/9404362}
\end{flushright}
\vspace{2.5cm}
{\LARGE
Model-Independent $Z'$ Limits\\ from  Electron-Electron Collisions}
}
\author{Debajyoti Choudhury, Frank Cuypers\\
{\small Max-Planck-Institut f\"ur Physik, Werner-Heisenberg-Institut,
D-80805 M\"unchen, Germany}\vspace{0.3cm}\\
{\small Emails: {\tt debchou,cuypers@iws166.mppmu.mpg.de}}
\vspace{0.5cm}\\
Arnd Leike\thanks{Supported by the German Federal Ministry for Research
and Technology under contract No.~05~6MU93P.}\\
{\small Sektion Physik der LMU M\"unchen, Theresienstr. 37, D-80333 M\"unchen,
 Germany}\vspace{0.3cm}\\
{\small Email: {\tt leike@cernvm.cern.ch}}}
\maketitle
\begin{abstract}
Model independent constraints on the mass of an extra neutral gauge boson and
its couplings to charged leptons are given for the $e^-e^-$ option
of a future \lc.
Analytic exclusion limits are derived in the Born approximation.
The results are compared with those of the $e^+e^-$ mode.
The influence of radiative corrections is discussed.
\end{abstract}
\thispagestyle{empty}
\newpage
%
%

Electron colliders
are usually assumed to be
electron-positron colliders.
However,
this need not necessarily be the case
at one of the projected \lc s
such as CLIC, JLC, TESLA, VLEPP, {\em etc.}.
Indeed,
a \lc\ can also be operated with two colliding electron beams.
Such an operation mode has two important advantages:
$(i)$ both beams can now be polarized to virtually 100\%;
$(ii)$ since QCD enters the game only at the two-loop level,
$e^-e^-$ collisions provide a very clean environment
for detecting slight deviations from the expectations
of the electro-weak sector of the \sm.
Of the many new possibilities \cite{phf,wo19,hm,cc},
not the least is a search for an extended electro-weak gauge sector.

Model independent constraints on an extra neutral gauge boson $Z'$
have been obtained previously for $e^+e^-$ collisions
in Refs~\cite{DESY93-154,delaguila}.
Here, we provide similar bounds for $e^-e^-$ collisions
and compare the results
with those obtained in $e^+e^-$ collisions.
For this we concentrate on a typical \lc\ design of the next generation,
assuming a \cm\ energy $\sqrt{s} = 500$ GeV
and an integrated luminosity ${\cal L} = 10$ fb$^{-1}$.
These values can anyway be modified trivially.
Since very high degrees of longitudinal polarization
should be available at future linear colliders,
we also assume 100\%\ polarization in the following.
The effects of dilution can be easily incorporated, though.

With the introduction of a $Z'$,
the Lagrangian for the relevant sector of the theory becomes
\begin{equation}
{\cal L} = e \left(\nobodyfrac
A_\mu J^\mu_\gamma + Z_\mu J^\mu_Z + Z'_\mu J^\mu_{Z'}\right),
\end{equation}
where $e$ is the electric charge
and $A_\mu$, $Z_\mu$ and $Z'_\mu$ represent the photon, the $Z$-boson
and the $Z'$.
The neutral currents are conveniently parametrized as
\begin{equation}
J^\mu_i = \bar\psi_e \gamma^\mu \left[  R_i P_R + L_i P_L \right] \psi_e
\qquad
(i=\gamma,Z,Z') \ ,
\end{equation}
where the left and right projection operators
are defined as $P_{R,L} \equiv (1 \pm \gamma_5)/2$.
The \sm\ left- and right-handed couplings $L_i$ and $R_i$
of the vector boson $i$ to electrons are
\begin{equation}
L_\gamma= R_\gamma=-1,\ \ \
L_Z=\tan\theta_W,\ \ \
R_Z= \tan\theta_W - \frac{1}{2\cos\theta_W\sin\theta_W},
\end{equation}
where $\theta_W$ is the electro-weak mixing angle.
The objective is now to obtain constraints on the
couplings $L_{Z'}$ and $R_{Z'}$ as a function of $m_{Z'}$.

At the Born level, the scattering
 $e^-(k_1)e^-(k_2) \rightarrow e^-(p_1)e^-(p_2)$
is described  by the exchange of neutral gauge bosons
in the $t$-- and/or $u$--channels,
depending on the polarization of the electron beams.
As the polarization of the final state electrons is as yet
impossible to determine experimentally,
the only observable is their  angular distribution.
We denote the cosine of this angle
\be
x = \cos\theta
\label{x}
\ee
and define
\be
   x_i = 1 + 2 \frac{m_i^2}{s} \qquad (i=\gamma,Z,Z') \ .
        \label{xi}
\ee
Neglecting the electron mass and the widths of the gauge bosons,
we have for the three possible combinations of beam polarizations
\be
\label{cross}
\barr{rcl}
 \displaystyle
   \frac{d\sigma^{LL}}{dx} &=  & \displaystyle \frac{16\pi\alpha^2}{s}
\sum_{i,j=\gamma,Z,Z'}L_i^2L_j^2
\left[ \frac{x_i x_j}{(x_i^2 - x^2) (x_j^2 - x^2) }  \right]
     \\[3.5ex]
 \displaystyle
   \frac{d\sigma^{RR}}{dx} &=  & \displaystyle \frac{16\pi\alpha^2}{s}
\sum_{i,j=\gamma,Z,Z'}R_i^2 R_j^2
\left[ \frac{x_i x_j}{(x_i^2 - x^2) (x_j^2 - x^2) }  \right]
     \\[3.5ex]
 \displaystyle
   \frac{d\sigma^{LR}}{dx} &=  & \displaystyle \frac{\pi\alpha^2}{s}
\sum_{i,j=\gamma,Z,Z'} L_i R_i L_j R_j
  \left[ \frac{(1 + x)^2}{ (x_i - x) (x_j - x)}
      + \frac{(1 - x)^2}{ (x_i + x) (x_j + x)} \right]
  \\[3.5ex]
 \displaystyle
   \frac{d\sigma^{ {\rm Unp.} } }{dx} &=  &
\displaystyle{1\over4} \left(
 \displaystyle
 \frac{d\sigma^{LL}}{dx} +  \frac{d\sigma^{RR}}{dx} +
      2  \frac{d\sigma^{LR}}{dx} \right) \ ,
\earr
\ee
where $\alpha = e^2/4\pi$ is the fine structure constant.
Note that since the polarization of the final state electrons
cannot be measured,
one has to sum over their polarizations.
This is why in the $LR$ case
the angular distribution remains symmetric.

As expected for M\o ller scattering,
the distribution becomes singular for $|x|\rightarrow 1$.
Had we retained the electron mass, it would have regulated
this collinear singularity arising from diagrams with  photon exchange.
This contribution is nonetheless eliminated naturally by the
experimental acceptance cut on the angle of the emergent electron,
$|x| < x_+$.

Before embarking on a more detailed analysis,
let us estimate the resolving power of this reaction
in the limit where
$m_{Z'} \gg \sqrt{s} \gg m_Z$,
hence neglecting terms of $O(M_Z^2/s)$ and $O(s/M_{Z'}^2)$.
The differences between the \xs s expected
in the presence and the absence of a $Z'$
become then
\begin{eqnarray}
\Delta\sigma_{LL}
&=&
L'^2 \qquad
{16\pi\alpha^2 \over s}\
\qquad {1 \over \cos^2\theta_W} \qquad\
\ln {1+x_+ \over 1-x_+}
\label{ll}
\\
\Delta\sigma_{RR}
&=&
R'^2 \qquad
{16\pi\alpha^2 \over s}\
{1 \over 4\sin^2\theta_W\cos^2\theta_W}\
\ln {1+x_+ \over 1-x_+}
\label{rr}
\\
\Delta\sigma_{LR}
&=&
L' R' \quad\ \
{4\pi\alpha^2 \over s}\ \
\qquad {1 \over \cos^2\theta_W} \quad\
\left( \ln {1+x_+ \over 1-x_+} - {3\over2}x_+ \right) \ ,
\label{lr}
\end{eqnarray}
where we defined the reduced left and right $Z'$ couplings
\begin{eqnarray}
L' &=& {\sqrt{s} \over m_{Z'}} L_{Z'}
\label{lp}
\\
R' &=& {\sqrt{s} \over m_{Z'}} R_{Z'} \ .
\label{rp}
\end{eqnarray}
Demanding that the difference in the number of events
is sufficiently significant,
leads to simple bounds on these reduced coupling.
Several features of the analysis are clear:
\begin{itemize}
\item	The dependence on the $Z'$ mass has been absorbed
	into the definition (\ref{lp},\ref{rp}) of the reduced couplings.
\item 	The $LL$ mode, with both beams left--polarized,
	is sensitive only to the coupling $L'$,
	and similarly for right polarization.
	There is thus no correlation
	between these two complementary measurements,
	which yield straight vertical and horizontal bands
	for the detectability limits in the $(L',R')$ plane.
\item	In contrast, the experiment with $LR$ beams
	yields highly correlated information on the $L'$ and $R'$ parameters.
	The curves delimiting the detectability region are now hyperbolas.
\item	In the limit where $\sin^2\theta_W = 1/4$
	the $LL$ and $RR$ modes have the same resolving power.
	In practice, the $RR$ mode yields only a minute improvement.
	The $LR$ mode, however, is much less sensitive.
\end{itemize}
All these features remain accurate
in the more precise analysis which is to follow now.

To take advantage of the angular information
contained in Eqs~(\ref{cross}),
we consider a moderate number $N$ of bins in $x=\cos\theta$
and compare the observed number of events $n_i$ in each
with the \sm\ expectations $n_i^{SM}$.
Denoting the fraction of events
in each bin by
\be
X_i = {n_i \over n}
\label{frac}
\ee
where $n = \displaystyle\sum_{i=1}^N n_i$,
a $\chi^2$ test for the
deviation can be devised as
\begin{equation}
\label{chi2}
\chi^2=
\sum_{i=1}^{N}\left(\frac{X_i-X^{SM}_i}{\Delta X_i}\right)^2 \ .
\end{equation}
The corresponding statistical error in the bin $i$ is given by
\be
\Delta X_i^2 = {n_i \over n^2} \left( 1-{n_i \over n} \right) \ .
\label{error}
\ee
The second term in Eq.~(\ref{error})
originates from the correlation between the number of events in one bin
and the total number of events.
The advantage of using the relative numbers of events $X_i = n_i/n$
resides in the fact that
the systematic error due to uncertainties in the luminosity measurements
drops out.

Armed with the above expressions, we can now examine the observabilty of
the reduced $Z'$ couplings (\ref{lp},\ref{rp}).
For concreteness we assume a \cm\ energy of 500 GeV and an integrated
luminosity of $10\, {\rm fb}^{-1}$. We impose the symmetrical angular cut
\be
|x| < x_+ = 0.985
  \label{cut}
\ee
and divide this range into $N=10$ equal size bins.
Since we deal here with one-sided bounds
the exclusion contours at 95\%\ confidence level in the $(L',R')$ plane
are identified by demanding $\chi^2 > 4.61$ in Eq.~(\ref{chi2}).
These contours are depicted for $m_{Z'}=2$ TeV
in Fig.~1 for $LL$, $RR$ and $LR$ beam polarizations.
These exclusion regions are indeed bounded by straight lines and hyperbolas,
as expected from the crude analysis based on Eqs~(\ref{ll}-\ref{lr}).
Obviously,
unpolarized beams are less sentive.
The combined fit
$\chi^2 = \chi^2_{LL} + \chi^2_{RR} + \chi^2_{LR}$
does not yield very much additional information,
since the two most sensitive measurements ($LL$ and $RR$)
already provide uncorrelated results.

While the results in Fig~1
are strictly speaking valid
only for the particular value 2 TeV of the $Z'$ mass,
the generic features prevail for other masses too.
Indeed,
the limits of detectability of $L'$ and $R'$
in the $LL$ and $RR$ experiments,
depend very little on the details of the analysis.
This can be infered from the plots
in Figs~2, 3 and 4,
where the smallest detectable value of $R'$
with 95\%\ confidence
(since the value taken by $L'$ is irrelevant,
this is only a one-parameter fit,
hence $\chi^2 = 2.71$),
is plotted respectively as a function of the cut $x_+$, 
the number of bins $N$,
and the mass $m_{Z'}$ of the $Z'$.
The reduced coupling $L'$ yields nearly undistinguishable results
(which would actually be identical for $\sin^2\theta_W=1/4$).

In Fig.~4 we have also plotted the discovery limits
which can be achieved by a $LR$ asymmetry measurement\footnote{
	This turns out to be the most sensitive experiment
	of Ref.~\cite{DESY93-154} to the variable $R'$}
in $e^+e^-$ collisions,
under the same conditions.
Clearly,
on the $Z'$ peak
(here at 500 GeV)
$e^+e^-$ collisions provide almost unlimited precision.
However,
if the $Z'$ mass exceeds the \cm\ energy by as little as 20\%,
the $e^-e^-$ mode with both beams polarized provides already more
accurate bounds.
Asymptotically,
roughly a factor of 1.6 can be achieved.

Finally, we want to comment on radiative corrections.
The considered reaction has
no resonating behaviour with the \cm\ energy $\sqrt{s}$. However, the cross
section has a very singular angular dependence keeping most of the outgoing
electrons at very small angles in the beam pipe.
The radiation of hard photons could kick an electron from the beam pipe
into the detector.
Nevertheless,
the corresponding hard photon can of course be vetoed.
The radiation of collinear hard photons from the initial state reduces the
effective energy of the colliding electrons giving a lower sensitivity
to a $Z'$.  However, such events can also be removed by demanding that the
energy of the scattered electrons be close to the beam energy.
To summarize, radiative corrections are not expected to induce
sizable changes to our model independent $Z'$ limits obtained at the Born
level.

To conclude,
we have demonstrated the excellent potential of $e^-e^-$ collisions in
constraining $Z'$ physics.
Model independent limits have been derived,
which are much more stringent than those
that could be obtained in  $e^+e^-$ collisions.
%
%
\ \vspace{1cm}\\ {\Large\bf Acknowledgement\hfill}\vspace{0.5cm}\\
One of us (A.L.) would like to thank T. Riemann and M. Bilenky for
usefull discussions.

\newpage
%

\newpage
{\bf\Large Figure Captions}\vspace{0.5cm}\\
\begin{description}
\item[Fig.~1]
Contours of Eq.~(\protect\ref{chi2}) for $\chi^2 = 4.61$
in the plane of the reduced $Z'$ couplings $(L',R')$
(\protect\ref{lp},\protect\ref{rp}),
for different combinations of beam polarizations.
The combined fit of the $LL$, $RR$ and $LR$ polarizations
is also shown.
\item[Fig.~2]
Dependence of the smallest observable value of the reduced coupling $R'$
(\protect\ref{rp})
with 95\%\ confidence,
as a function of the angular cut (\protect\ref{cut}).
\item[Fig.~3]
Dependence of the smallest observable value of the reduced coupling $R'$
(\protect\ref{rp})
with 95\%\ confidence,
as a function of the number of bins in Eq.~(\protect\ref{chi2}).
\item[Fig.~4]
Dependence of the smallest observable value of the reduced coupling $R'$
(\protect\ref{rp})
with 95\%\ confidence,
as a function of the mass of the $Z'$.
\end{description}

\begin{thebibliography}{99}
\bibitem{phf} P.H.~Frampton and D.~Ng, {\em Phys.~Rev.} {\bf D 45} (1992) 4240.
\bibitem{wo19} F.~Cuypers, G.J.~van Oldenborgh and R.~R\"uckl,
{\em Nucl.~Phys.} {\bf B 409} (1992) 128.
\bibitem{hm} C.A.~Heusch and P.~Minkowski, CERN preprint CERN-TH-6606-92,
May 1993.
\bibitem{cc} D.~Choudhury and F.~Cuypers, MPI preprint MPI-Ph/94-24.
\bibitem{DESY93-154} A. Leike, DESY 93-154, to appear in Z. Phys. {\bf C}.
\bibitem{delaguila} F. del Aguila, UG-FT-35/94.
\bibitem{zpsari} A. Djouadi, A. Leike, T. Riemann, D. Schaile, C. Verzegnassi,
                    Proc. of the ``Workshop on Physics and Experiments with
                    Linear Colliders'', Sept. 1991, Saariselk\"a, Finland, ed.
                    R. Orava, Vol. II, p. 515;\\
                 A. Djouadi, A. Leike, T. Riemann, D. Schaile, C. Verzegnassi,
                    Z. Phys. {\bf C56} (1992) 289.
\end{thebibliography}
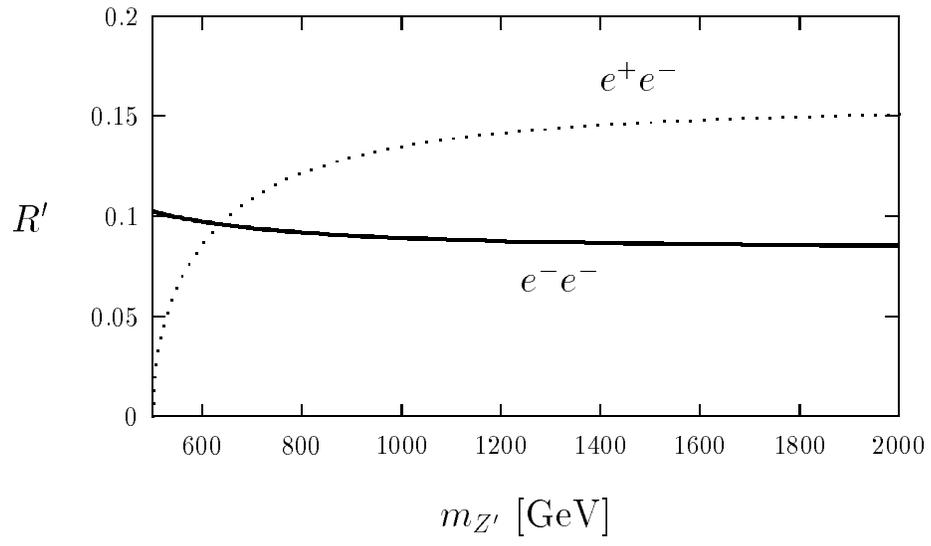
\end{document}